\documentclass[prl,twocolumn,superscriptaddress,showpacs]{revtex4}
\usepackage{amsmath,amssymb,graphicx,subfigure,color}

\newcommand{\scs}{\scriptscriptstyle}

\def\a{\alpha}

\def\b{\beta}

\def\<{\langle}
\def\>{\rangle}
\begin{document}
\title{The Glass-like Structure of Globular Proteins and the Boson Peak}
\author{Stefano Ciliberti}
\affiliation{Laboratoire de Physique Th\'eorique et Mod\`eles Statistiques,
Universit\'e de Paris-Sud, b\^atiment 100, 91405, Orsay Cedex, France}
\author{Paolo De Los Rios} 
\author{Francesco Piazza} 
\affiliation{Laboratoire de Biophysique Statistique-ITP, EPFL, CH-1015 Lausanne, Switzerland}
\date{\today}
\begin{abstract}
  Vibrational spectra of proteins and topologically disordered solids display
  a common anomaly at low frequencies, known as Boson peak. 
  We show that such feature in  globular proteins can be deciphered in terms of an 
  energy landscape picture, as it is for glassy systems. Exploiting the tools of Euclidean
  random matrix theory, we clarify the physical origin of such
  anomaly  in terms of a mechanical instability of the system.
  As a natural explanation, we argue 
  that such instability is relevant for proteins in order for their molecular 
  functions to be optimally rooted in their structures.
 

  \end{abstract}
%
\pacs{87.15.-v,87.15.He;63.50.+x,64.70.Pf}
%
%
%
%
%
\maketitle

%
%
Proteins are characterized by mechanically stable, unique native structures
that bear a precise relation with their biological functions. Yet, in most
cases, specific functionality is accompanied by large-amplitude dynamical
conformational changes that require high
flexibility~\cite{frauenfelder}. Protein structures are complex,
hierarchical ones, characterized by short-range order and overall spatial
correlations that bear strong similarities with those of glassy
materials~\cite{disorder}.  In actual fact, proteins and glasses share many
physical properties, such as peculiar relaxation processes~\cite{noiPRL} and the
occurrence of a dynamical transition as revealed by the temperature
dependence of the atomic mean square displacements
(MSD)~\cite{frauenfelder,prot-glass1,prot-glass2}.

Interestingly, there exists a remarkable similarity of the Raman and
neutron--scattering spectra of proteins with those of glasses and
super-cooled liquids~\cite{prot-glass1}, i.e. a peak that develops at low
temperatures in the low-frequency regions.  Such anomaly, known as Boson
peak (BP), also shows up in the experimentally determined density of states
when divided by the Debye law, i.e. $g(\omega)/\omega^2$~\cite{bpprotexp}.
Several models have been proposed for the explanation of the BP in proteins,
among which the phonon-fracton model~\cite{phon-fracton}, and the log-normal
distribution model~\cite{log-normal}.

The BP is, on the other hand, a universal feature of many glassy
systems~\cite{BPexp}.  In this context, several possible explanations have
been proposed, from the two-level system scenario~\cite{phillips87} to localized
modes arising from a strong scattering of the phonons by the
disorder~\cite{foret96}, from ``glassy'' van Hove
singularities~\cite{taraskin01} to a mechanical instability~\cite{parisi02}.
Recently, the possibility that a BP may be a general feature of weakly connected
systems has also been investigated~\cite{wyart04,liu05}. 

In a different analytical framework~\cite{ciliberti03}, 
the excess of low-energy modes with respect to the Debye behaviour is viewed as a
symptomatic effect of the topological phase transition which is conjectured
to happen in glasses at low temperatures~\cite{parisi02}. Recently, a
quantitative description of the BP phenomenology has been given whithin the
formalism of the Euclidean Random Matrix (ERM)
theory~\cite{ciliberti03}, whose predictions have been confirmed by numerical simulations 
on realistic glass-forming systems, emphasizing its universal character~\cite{grigera03}.

In this Letter, we show that the emergence of a BP in globular proteins is
the signature of a structural instability of the saddle-phonon kind akin to
that predicted within the ERM theory of glasses.  Remarkably, our
explanation allows for a natural interpretation of such instability in
proteins in terms of the mutual relations among their structure, dynamics
and biological function.

%
%
%
%
To investigate the vibrational properties of a given globular protein, we
coarse-grain its structure at the amino-acid level and build the associated
elastic network (EN).  The application of EN models to proteins is relatively
recent~\cite{Tirion}, since it has commonly been assumed that little
structural detail could be given up in order to model their complex energy
landscapes.  However, there is now strong evidence that most features of the
large- and medium-scale dynamics of proteins' fluctuations around their
native states, related to function and stability,  
can be successfully reproduced by simple harmonic interactions
between amino-acids~\cite{atilgan01, Bahar, hinsen, yhs,turco}.  In view of the BP phenomenology,
it is important to mention the growing consensus that an explanation 
in glasses could be found within a purely harmonic context~\cite{harmonic}.

In the framework of EN models, the potential energy is written as a sum of pair-wise
harmonic potentials,
\begin{equation}
\label{e:totpot}
\mathcal{V}(\{ \vec{r}\}) = \sum_{i<j} V(\vec{r}_i,\vec{r}_j) \\
                =  
                    \sum_{i<j} \frac{k_{ij}}{2} 
                         \left(
                          |\vec{r}_{ij}|-|\vec{r}^{\, \scs (0)}_{ij}|
                         \right)^2
\end{equation}
where $\vec{r}_{ij}=\vec{r}_i-\vec{r}_j$, $\vec{r}_i$ being the position of
the $i$--th particle, $\vec{r}^{\, \scs (0)}_{i}$ its equilibrium position and
$k_{ij}$ the stiffness of the spring connecting particles $i$ and $j$.
More precisely, the vector $\vec{r}_i$ represents the instantaneous position
of the $\alpha$-carbon of the $i$-th amino-acid, $\vec{r}^{\, \scs (0)}_{i}$ its
position in the native state as determined from X-ray crystallography or
Nuclear Magnetic Resonance, and $k_{ij}$ can take different functional
forms, such as $k_{ij}= \kappa \theta(r_c-|\vec{r}_{i}^{\, \scs (0)}-\vec{r}_{j}^{\, \scs (0)}|)$
(sharp cutoff model~\cite{Bahar}) or $k_{ij}=\kappa
\exp(-|\vec{r}_{i}^{\, \scs (0)}-\vec{r}_{j}^{\, \scs (0)}|^2/r_c^2)$ (Gaussian model~\cite{hinsen}),
which is the one we adopt here.  The parameter $\kappa$ sets the physical
units for force constants, and can be fixed by requiring the theoretical
MSDs to match the experimental ones as determined from X-ray
spectra~\cite{atilgan01}.  

\begin{figure}[t!]
\centering
\includegraphics[width=\columnwidth,clip]{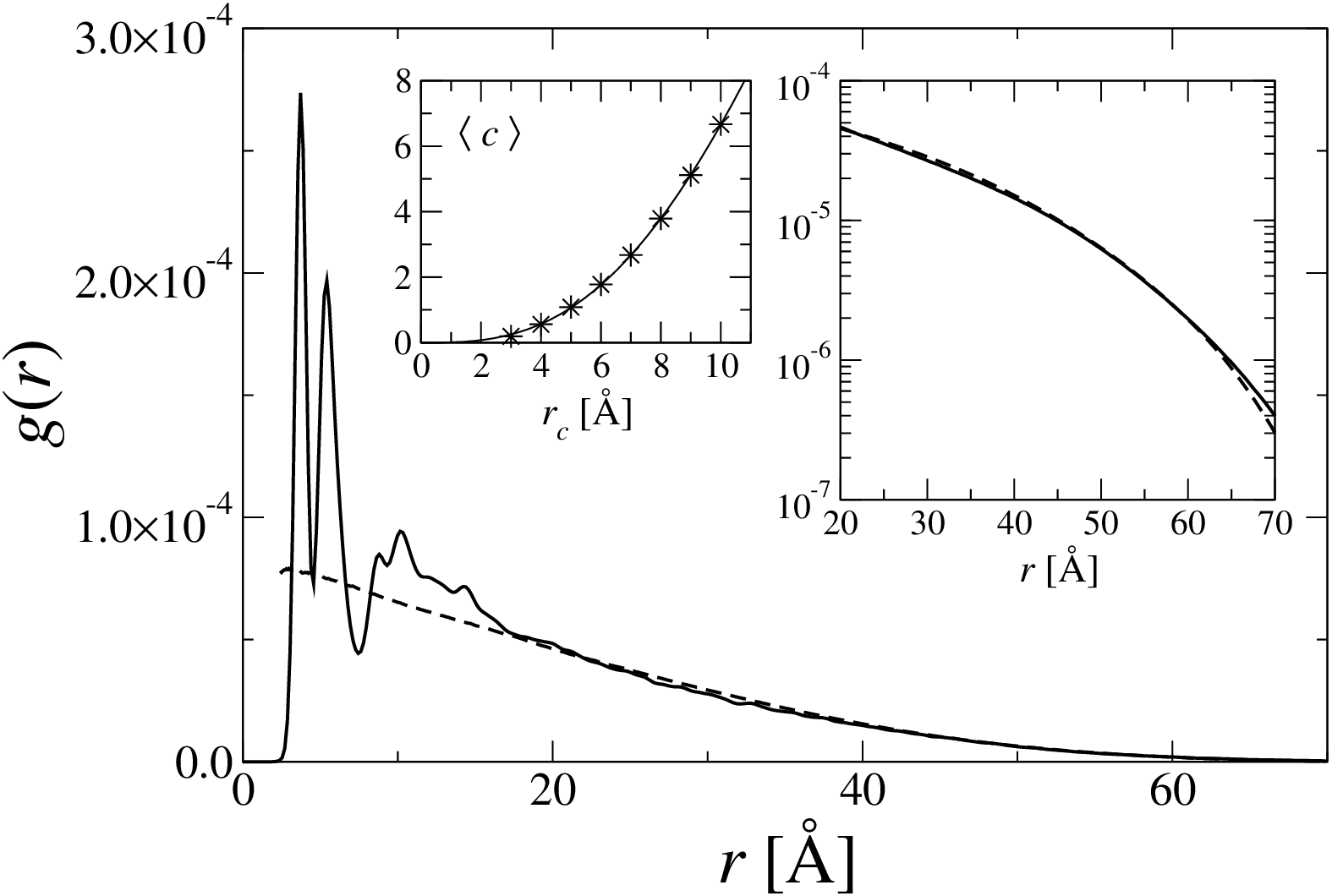}
\caption{Plot of the pair correlation function for Serum Albumin ($N=578$,
solid line) and for a collection of an equal number of residues uniformly
distributed in its equivalent ellipsoid (dashed line). 
Right inset: a magnification of the tails in lin-log scale.
Left inset: Average connectivity vs cutoff distance (symbols) 
and cubic fit (solid line).}
\label{f:Gdir}
\end{figure}

In the harmonic approximation, the total
potential energy~(\ref{e:totpot}) is the quadratic form $\mathcal{V}(\{
\vec{r}\}) = \frac{1}{2} \vec{r}^{\, \scs T} \mathbb{K} \, \vec{r}$, where
the contact matrix $\mathbb{K}_{i\alpha,j\beta}$ ($\alpha,\beta=1,2,3$) 
is the Hessian of the function~(\ref{e:totpot}) evaluated at the equilibrium structure.
Were the position vectors in the native structure $\vec{r}_{i}^{\, \scs (0)}$ arranged at
random, $\mathbb{K}$ would exactly fall in the class of Euclidean Random
Matrices.  Even if protein structures are surely not random, an
analysis of the pair correlation function $g(r)$ reveals interesting
features.  In Fig.~\ref{f:Gdir} we plot $g(r)$ for Serum Albumin, a
relatively large globular protein whose equivalent ellipsoid~\cite{sheraga}
has principal radii measuring 2.3, 3.7 and 4 nm, and for an identical number
of residues uniformly distributed within such ellipsoid. The comparison shows
that the protein structure is characterized by two well-defined
coordination shells, namely the nearest neighbors at fixed distance along the
chain and the next-nearest off-chain neighbors, including the pairs belonging 
to alpha helices and those lying at turning regions, such as loops.
After a third, less resolved shell all pair-wise spatial correlations are lost.  We repeated this
analysis for several proteins and always found  that the
second and the third peaks are always related to the presence of secondary
motifs as well as to the intrinsic flexibility of the peptide chain, while beyond such
range spatial correlations are absent.
This fact is a clear indication that, as far as large-scale structural properties are
involved, proteins are well approximated by random assemblies of amino-acids 
with specified density. 

%
%
%
The analogy between protein structures and disordered systems with no
long-range order suggests that a common mechanism might be responsible for
the emergence of the BP in both cases.  In topologically disordered solids, 
this anomaly appears upon increasing the temperature or, as observed for example in Silica, 
upon lowering the
density. In the present case, we are dealing with proteins, i.e. objects
whose equilibrium structure is fixed by the biological function. 
However, changes in the particle density may still
be simulated by resorting to the free parameter
$r_{c}$.  In the framework of EN models, $r_{c}$ sets the range of
inter-particle interactions and should in principle be tuned 
by fitting the low-frequency portion of experimental spectra at temperatures below the dynamical
transition, where the protein vibrates harmonically within a local minimum.
The usual alternative is to compare with spectra
as determined by  all-atom force fields~\cite{hinsen}. 
By doing this, one obtains $\rho_{c} \approx 3$ \AA \ in
an all-atom representation~\cite{hinsen}, which coarse-grains to 
$r_{c} \approx \< N_{a} \>^{1/3}\rho_{c} \approx 8$ \AA \ when the average number of
atoms per amino-acid $\< N_{a} \> \approx 18$ is introduced.
Interestingly, by its very definition, the parameter $r_{c}$ also allows to
regulate an effective {\em local density} of the system by tuning the
average connectivity  $\langle c \rangle \equiv \frac{1}{3N}
\sum_{i=1}^N \sum_{\alpha=1}^3\mathbb{K}_{i\alpha,i\alpha}$.
%
%
By decreasing the cutoff $r_{c}$, the average number of neighbors per
residue diminishes accordingly. Thus, a local measure of compactness may be
introduced that is proportional to $\langle c \rangle$. It can be shown that
varying $r_{c}$ induces a change in the connectivity that scales
with the interaction volume $r_{c}^3$ up to finite-size $\mathcal{O}(r_{c})$
corrections (see left inset in Fig.~\ref{f:Gdir}).  This means that we can study the 
spectral features of a given protein structure with the additional degree of freedom 
of varying density by simply changing the interaction cutoff $r_{c}$, which thus plays in this
context the role of a control parameter.

The vibrational spectrum of a protein for a certain value of the parameter
$r_{c}$ is obtained by diagonalizing the contact matrix.  However,
especially for small proteins, the finite number of residues makes it
difficult to analyze the low-frequency features of the spectra.  In order to
circumvent this problem, we generated a number of different conformers for
each of the analyzed structures such that all of them be by construction
compatible with the atomic MSDs as specified by the native contact matrices.
More precisely, if we write the coordinates of a given conformer as
$\vec{\rho}^{\, \scs (0)} = \vec{r}^{\, \scs (0)} + \delta \vec{r}$, then it is sufficient to
take $\delta \vec{r} = \mathbb{U}
\, \vec{c}$, where $\mathbb{U}$ is the matrix of eigenvectors of
$\mathbb{K}$ and the $3N-6$ coefficients $c_{k}$ are drawn from as many one-dimensional
Gaussian distributions with zero mean and standard deviations
$\sigma_{k}=\sqrt{-k_{\scs B}T/\lambda_{k}}$, $\lambda_{k}=-\omega_{k}^2$
being the eigenvalues of the contact matrix $\mathbb{K}$. 
This procedure provides a simple means to construct an arbitrary number of
conformations that are dynamically equivalent to the native one in the
harmonic approximation.

\begin{figure}[t!]
\centering
\subfigure{
\includegraphics[width=\columnwidth,clip]{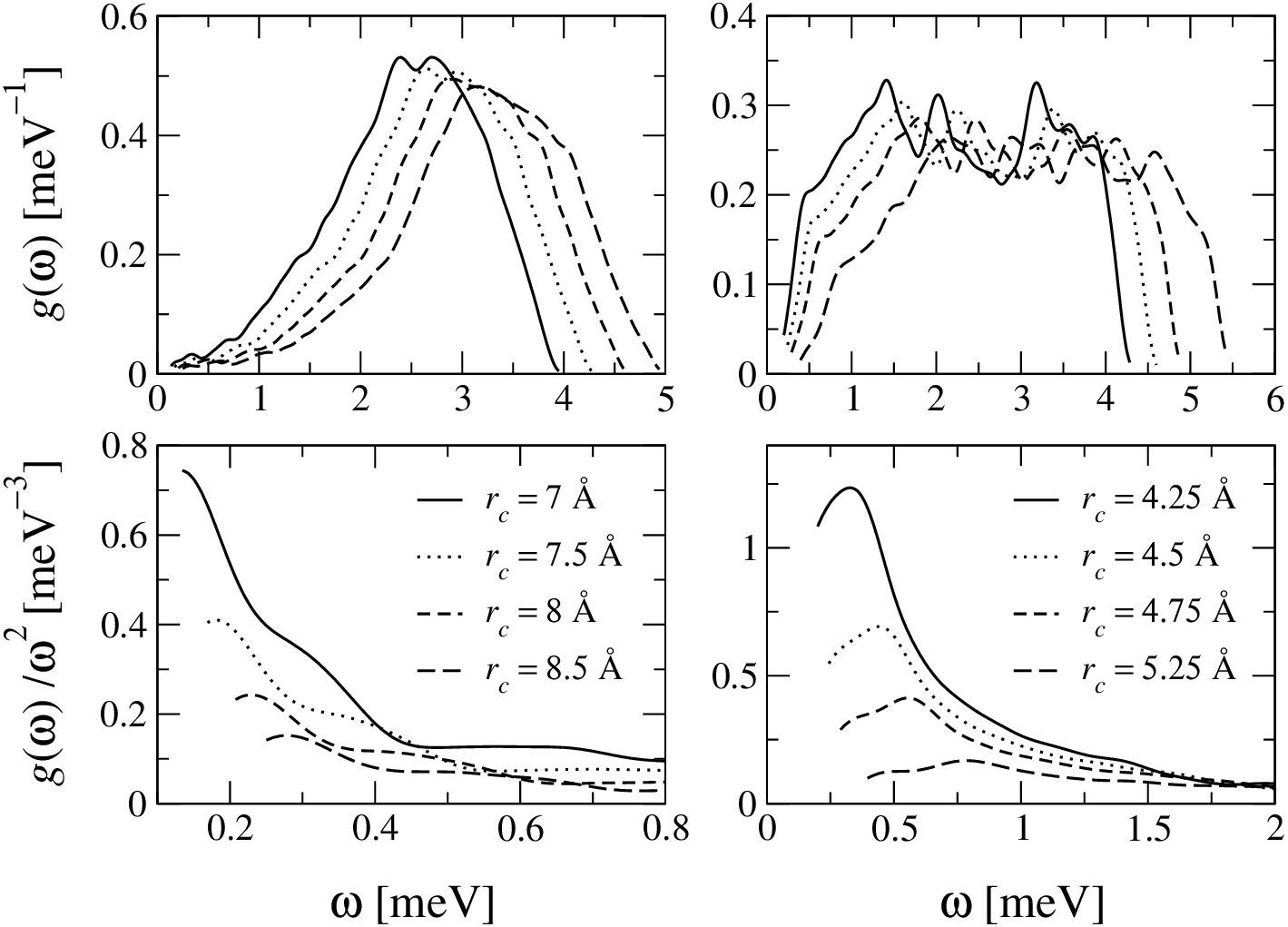}}
\subfigure{
\includegraphics[width=\columnwidth,clip]{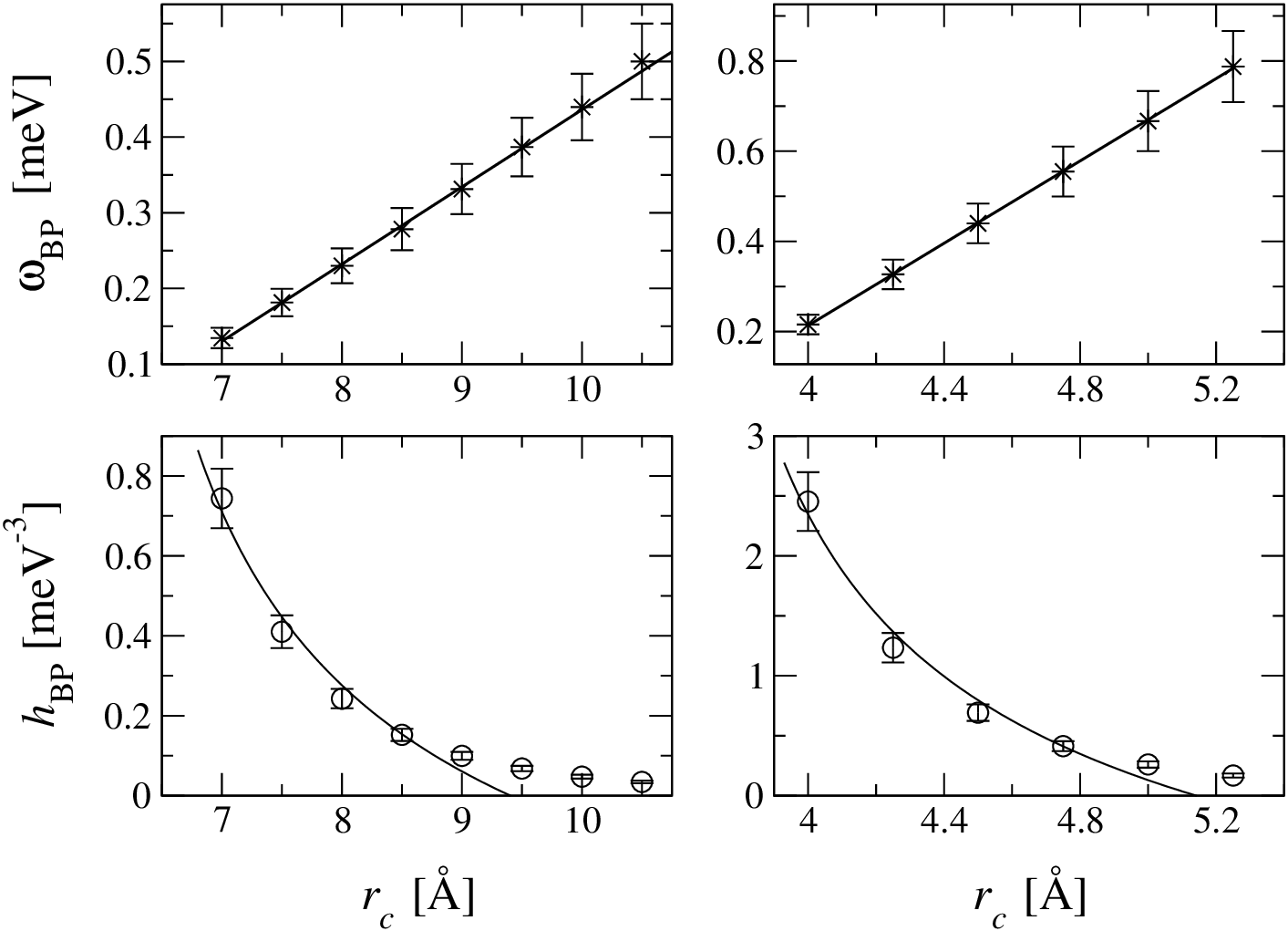}}
\caption{Boson peak analysis for two globular proteins of different size.
Left panels:  Serum Albumin (1AO6), $N=578$ residues.  Right panels:
Ubiquitin (1UBI), $N=76$ residues.  The four upper panels show the density
of states for different values of $r_{c}$ (for 1000 thermal replicas).  In
the four lower ones, we show the fits to the BP frequency and height with
the mean field expressions~(\ref{e:BP-ERM}).  The best-fit results are:
$r_c^{\scs \ast}=5.7$ \AA \ (Serum Albumin) and $r_c^{\scs \ast}=3.5$ \AA \ (Ubiquitin).
The physical units for frequencies were obtained with $r_{c}=8$ \AA.}
\label{f:BP}
\end{figure}

In Fig.~\ref{f:BP} we plot $g(\omega)$ and $g(\omega)/\omega^2$ for several values of 
the cutoff $r_{c}$ for two representative proteins of different size.  Similar
results were obtained for a choice of other proteins.
A shoulder manifestly  appears in the low-frequency region as $r_{c}$ is reduced 
(see upper panels in Fig~\ref{f:BP}), and eventually a divergence develops if $r_{c}$ is
decreased below a critical value.  The origin of such peak can be uncovered
by tracking its position $\omega_{\rm BP}$ and height $h_{\rm BP}$ as
$r_{c}$, i.e. our effective density, decreases. From the lower panels of
Fig~\ref{f:BP} one can clearly appreciate that the scaling followed by
$\omega_{\rm BP}$ and $h_{\rm BP}$ is very well interpolated by the
analytical functional forms predicted by the ERM theory in the mean-field
approximation~\cite{ciliberti03}, i.e.
\begin{equation}
\label{e:BP-ERM}
\omega_{\rm BP} \sim (r_{c}-r_c^{\scs \ast})^\alpha,
\quadÊ h_{\rm BP}      \sim ({r_{c}-r_c^{\scs \ast}})^{-\beta}
\end{equation}
with $\alpha=1$ and $\beta=1/2$. Therefore, our analysis strongly suggests
that the BP in protein structures at low densities can be interpreted in
terms of a topological instability utterly analogous to the one
found in glasses and glass-forming liquids~\cite{grigera03}.
More rigorously, as it is the case for the Gaussian model in glasses, the BP
should be interpreted as a precursor of the transition within a model that
by definition becomes meaningless at the critical point. This is precisely
what happens in our case, at an interaction range  below which protein structures
start unfolding. 
We also stress that the shift of $\omega_{\textrm {BP}}$ towards zero
frequency and the divergence of the BP height as the systems loose rigidity
is a spectral feature equally unveiled within different theoretical 
approaches~\cite{taraskin01,parisi02,wyart04,liu05}.

\begin{figure}[t!]
\centering
\includegraphics[width=\columnwidth,clip]{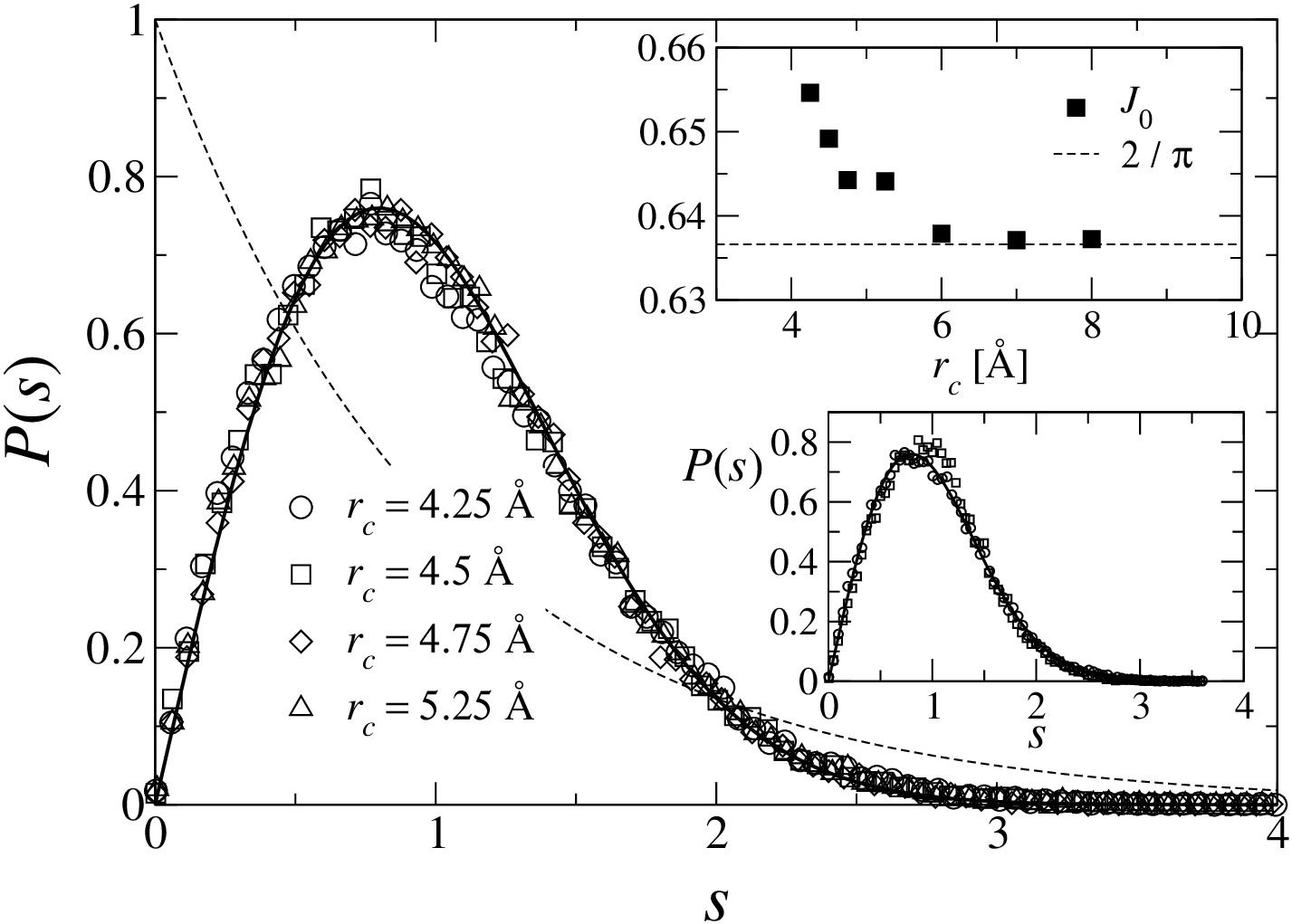}
\caption{Plot of the level spacing statistics of Ubiquitin for
different values of the cutoff $r_{c}$. The Wigner-Dyson (thick solid line)
and Poisson (dashed line) statistics, which describes totally uncorrelated
spectra, are also shown for comparison. Upper inset: $J_0\equiv\langle
s^2\rangle/2$ is plotted versus $r_{c}$. The dashed line represents the
value expected for a fully extended spectrum. Lower inset: level spacing
statistics for frequencies $\omega < 2.5 $ meV. The solid line is a plot of
the Wigner surmise.}
\label{f:loca1}
\end{figure}

It is also instructive to study the
localization properties of typical ensembles of spectra through the
level-spacing statistics  $P(s)$~\cite{guhr98}.
As an example, we plot the results obtained for Ubiquitin in
Fig.~\ref{f:loca1}. Overall, the distribution is very well described by a
Wigner law, which holds for fully extended spectra. As we decrease the
cutoff, a small contribution from localized modes is observed, as the
measure of $J_0\equiv \langle s^2\rangle/2$ shows (upper inset of
Fig.~\ref{f:loca1}). Otherwise, $J_0$ should be close to $1$ in the
case of a localized spectrum, which is never the case. A more refined
analysis~\footnote{see e.g. S. Ciliberti and
T. S. Grigera, Phys. Rev. E {\bf 70}, 061502 (2004).}  performed on several
proteins clearly shows that the only localized modes are due to the tail of the spectrum 
at large frequencies, much alike structural glasses~\cite{taraskin01,chumakov04}.
This conclusion, further confirmed by the level spacing statistics 
from the low-frequency portion of the spectra (lower inset of Fig.~\ref{f:loca1}),
rules out the presence of localized modes in the BP region.  

The origin of a precursory feature of a topological instability in proteins
can be formally understood by recalling that their structures are those of
folded polymers.  If the interaction cutoff $r_{c}$ is lowered below
the first off-chain coordination shell, native conformations lose their folded nature and become more and more akin 
to liquids. In fact, we argue that the appearance of the BP precisely anticipates such
inherent instability before the critical cutoff is reached.
Accordingly, the  best-fit values of $r_c^{\scs \ast}$ for all
the analyzed structures does never exceed the first off-chain coordination
shell (see Fig.~\ref{f:BP}).
Keeping in mind that the optimal value of $r_{c}$ is around $8$ \AA, i.e. above
its critical value, our results suggest that protein structures express an inherent trade off 
between spatial properties of liquids, i.e. increased degree of mobility,
and the necessity of maintaining a certain structural stability.  
Interestingly,  from an extensive analysis on a selection of 13 proteins,
we find that $r_c^{\scs \ast}$ is substantially anti-correlated
with the packing fraction $p = 4/3(N/V)(d_{0}/2)^3$, i.e. a measure of global compactness, whereas 
weak correlation is found with indicators of local stability, 
such as  the content of $\alpha$ helices and $\beta$ sheets.
Here $N$ and $V$ are the number of residues and the volume,
while $d_{0} \simeq 3.83$ \AA \ is the inter-residue distance along the main chain.
Moreover, we also find  a positive correlation between $r^\ast_c$  and $N$, which 
may signal the larger mechanical stability of smaller proteins (see Table~\ref{tab1}).

%
\begin{table}[t!]
\centering
\begin{tabular}{l||c|c|c||c}
\hline
{\bf Protein} & \quad $N$ \quad   &  Ê\quadÊ$p$Ê \quadÊ &  $(\a+\b)$ & $r_c^{\scs \ast}$ (\AA)\\
\hline\hline 
Insulin                            &        51          &        0.20         &      0.53         &   4.57  \\
Protein G                          &        56          &        0.21         &      0.70         &   3.64  \\
Ubiquitin                          &        71          &        0.20         &      0.46         &   3.53  \\ %
PDZ binding domain                 &        85          &        0.21         &      0.55         &   4.03  \\
Lysozyme                           &       162          &        0.17         &      0.74         &   4.27  \\
Adenylate Kinase                   &       214          &        0.12         &      0.64         &   7.85  \\ %
LAO                                &       238          &        0.16         &      0.60         &   5.44  \\
CYSB                               &       260          &        0.17         &      0.59         &   4.70  \\ %
PBGD                               &       296          &        0.16         &      0.60         &   3.70  \\ %
Thermolysin                        &       316          &        0.18         &      0.53         &   4.55  \\
HSP70 ATP-binding domain           &       382          &        0.15         &      0.66         &   5.28  \\
Fab-fragment                       &       437          &        0.13         &      0.48         &   5.70  \\ 
Serum Albumin                      &       578          &        0.12         &      0.70         &   5.70  \\ \hline\hline
Correlation with $r_c^{\scs \ast}$ &  $\mathbf{0.45}$   &  $\mathbf{ -0.82}$  &  $\mathbf{0.17}$  &   1      \\\hline
\end{tabular}
\caption{Correlation of $r_c^{\scs \ast}$ with structural parameters.}
\label{tab1}
\end{table}%

The above conclusions may be interpreted by regarding proteins as molecular machines
bound to keep a specified geometry in order to perform their biological function, yet
preserving a high degree of structural flexibility in order to 
efficiently explore different conformational states.
In this sense, the mechanical instability underlying the emergence of a BP appears to 
be a universal signature of their engineered ability to easily travel between 
adjacent local minima in their native states.
We note that our results agree with recent estimates of the spectral dimension
of globular proteins, whose non-Debye behavior has been interpreted
in terms of a vibrational instability of the Peierls-Landau type~\cite{cecconiscienza}.

Summarizing, in this Letter we have provided compelling evidence of the
equivalence of the Boson peak phenomenon in globular proteins and glasses.
Our analysis suggests that a topological instability of the
saddle-phonon type in proteins reflects the balance imprinted in 
their structures between being able of rapidly accessing different minima in
the native energy landscape while keeping a relative mechanical rigidity.

We thank L. Casetti, O. Martin, M. M\'ezard and G. Parisi for interesting discussions. 
S.~C. also thanks  the EPFL for its hospitality. S.~C. is supported by ECHP 
Program, contract HPRN-CT-2002-00319 (STIPCO).


\end{document}